\documentclass[final,5p,times,twocolumn]{elsarticle}

\usepackage{graphicx}
\usepackage{amssymb}
\usepackage{amsmath}
\usepackage{empheq} 
\usepackage{comment}
\usepackage{balance}
\usepackage{enumerate}
\usepackage[inline,shortlabels]{enumitem}

\biboptions{sort&compress} % cite as [1,2,3] and not [3,1,2] etc. 

\newcommand{\dd}{\mathrm{d}} % differential d

\makeatletter
\def\@author#1{\g@addto@macro\elsauthors{\normalsize%
    \def\baselinestretch{1}%
    \upshape\authorsep#1\unskip\textsuperscript{%
      \ifx\@fnmark\@empty\else\unskip\sep\@fnmark\let\sep=,\fi
      \ifx\@corref\@empty\else\unskip\sep\@corref\let\sep=,\fi
      }%
    \def\authorsep{\unskip,\space}%
    \global\let\@fnmark\@empty
    \global\let\@corref\@empty  %% Added
    \global\let\sep\@empty}%
    \@eadauthor={#1}
}
\makeatother

\makeatletter
\def\ps@pprintTitle{%
 \let\@oddhead\@empty 
 \let\@evenhead\@empty
 \def\@oddfoot{}%
 \let\@evenfoot\@oddfoot}
 \makeatother

\usepackage[table,xcdraw,svgnames]{xcolor}
\usepackage[colorlinks]{hyperref}
\AtBeginDocument{%
  \hypersetup{
    citecolor=SteelBlue,
    linkcolor=SteelBlue,   
    urlcolor=SteelBlue}
   }

\journal{~}
\begin{document}
\begin{frontmatter}

\title{GDP growth rates as confined L\'evy flights: \\
towards a unifying macro theory of economic growth rate fluctuations 
}

\author[add1,add2]{Sandro Claudio Lera}
\ead{slera@ethz.ch}

\author[add2,add3]{Didier Sornette}
\ead{dsornette@ethz.ch}

\address[add1]{\scriptsize ETH Zurich, Singapore-ETH Centre, 1 CREATE Way, \#06-01 CREATE Tower, 138602 Singapore}
\address[add2]{\scriptsize ETH Zurich, Department of Management, Technology, and Economics, Scheuchzerstrasse 7, 8092 Zurich, Switzerland}
\address[add3]{\scriptsize Swiss Finance Institute, c/o University of Geneva, Geneva, Switzerland}

% abstract
%%%%%%%%%%%%%%%%%%%%%%%%%%%%%%%%%%%%%%%%%%%%%%%%%%%%%%%%%%%%%%%%%%%%%%%%%%%%%%
\begin{abstract}

A new model that combines economic growth rate fluctuations at the microscopic and macroscopic level is presented. 
At the microscopic level, firms are growing at different rates while also being exposed to idiosyncratic 
shocks at the firm and sector level. 
We describe such fluctuations as independent L\'evy-stable fluctuations, varying over multiple orders of magnitude. 
These fluctuations are aggregated and measured at the macroscopic level in averaged economic output quantities such as GDP. 
A fundamental question is thereby to what extend individual firm size fluctuations can have a noticeable impact on the overall economy. 
We argue that this question can be answered by considering the L\'evy fluctuations as embedded in a steep confining potential well,  ensuring nonlinear mean-reversal behavior, without having to rely on microscopic details of the system. 
The steepness of the potential well directly controls the extend towards which idiosyncratic shocks to firms and sectors are damped at the level of the economy. 
Additionally, the theory naturally accounts for business cycles, represented in terms of a bimodal economic output distribution, and thus connects two so far unrelated fields in economics. 
By analyzing 200 years of US GDP growth rates, we find that the model is in good agreement with the data. 

\end{abstract}
%%%%%%%%%%%%%%%%%%%%%%%%%%%%%%%%%%%%%%%%%%%%%%%%%%%%%%%%%%%%%%%%%%%%%%%%%%%%%%

% keywords 
%%%%%%%%%%%%%%%%%%%%%%%%%%%%%%%%%%%%%%%%%%%%%%%%%%%%%%%%%%%%%%%%%%%%%%%%%%%%%%
\end{frontmatter}

\section*{Introduction}
\label{sec:intro}
%%%%%%%%%%%%%%%%%%%%%%%%%%%%%%%%%%%%%%%%%%%%%%%%%%%%%%%%%%%%%%%%%%%%%%%%%%%%%%

As a result of the increasing interconnectedness of firms, and more general, of society \cite{Schweitzeretal09}, the study of firm size fluctuations and its effect on the entire economy
has become an active area of research \cite{Gabaix2011,Acemoglu2012,diGiovanni2014,Friberg2016,Anthonisen2016,Stella2015}. 
A central question is how the productivity fluctuations of individual firms aggregate into productivity measures for the economy as a whole. 
Albeit still subject to debate \cite{Stella2015}, there is now a dominant view \cite{Gabaix2011,Acemoglu2012,diGiovanni2014,Friberg2016,Anthonisen2016}, that idiosyncratic shocks to individual large firms can have a significant effect on the macroeconomic output growth.
Explanations for this effect range from statistical fluctuation arguments \cite{Gabaix2011} to more intricate models taking into consideration intersectoral input-output linkages \cite{Acemoglu2012}
or dynamic income-expenditure networks \cite{Anthonisen2016}. 

In this article, we offer a different, more coarse-grained perspective. 
We show how heavy-tailed idiosyncratic shocks are mitigated through a kind of renormalized economic potential to the aggregate output growth distribution.
Specifically, we claim that the aggregate output growth rate distribution can be modeled as a L\'evy flight in a steep 
confining potential.
This establishes a quantitative connection between the tail distribution of idosyncratic shocks to individual firms and the aggregate economic growth distribution. 
To provide empirical support for our claim, we study the gross domestic product (GDP) of the United States over that last 200 years, and find that the model
is in good agreement with the data.

\section*{Structure of the GDP growth rate distribution}
\label{sec:GDP}
%%%%%%%%%%%%%%%%%%%%%%%%%%%%%%%%%%%%%%%%%%%%%%%%%%%%%%%%%%%%%%%%%%%%%%%%%%%%%%

\begin{figure*}[!htb]
	\centering
	\includegraphics[width=\textwidth]{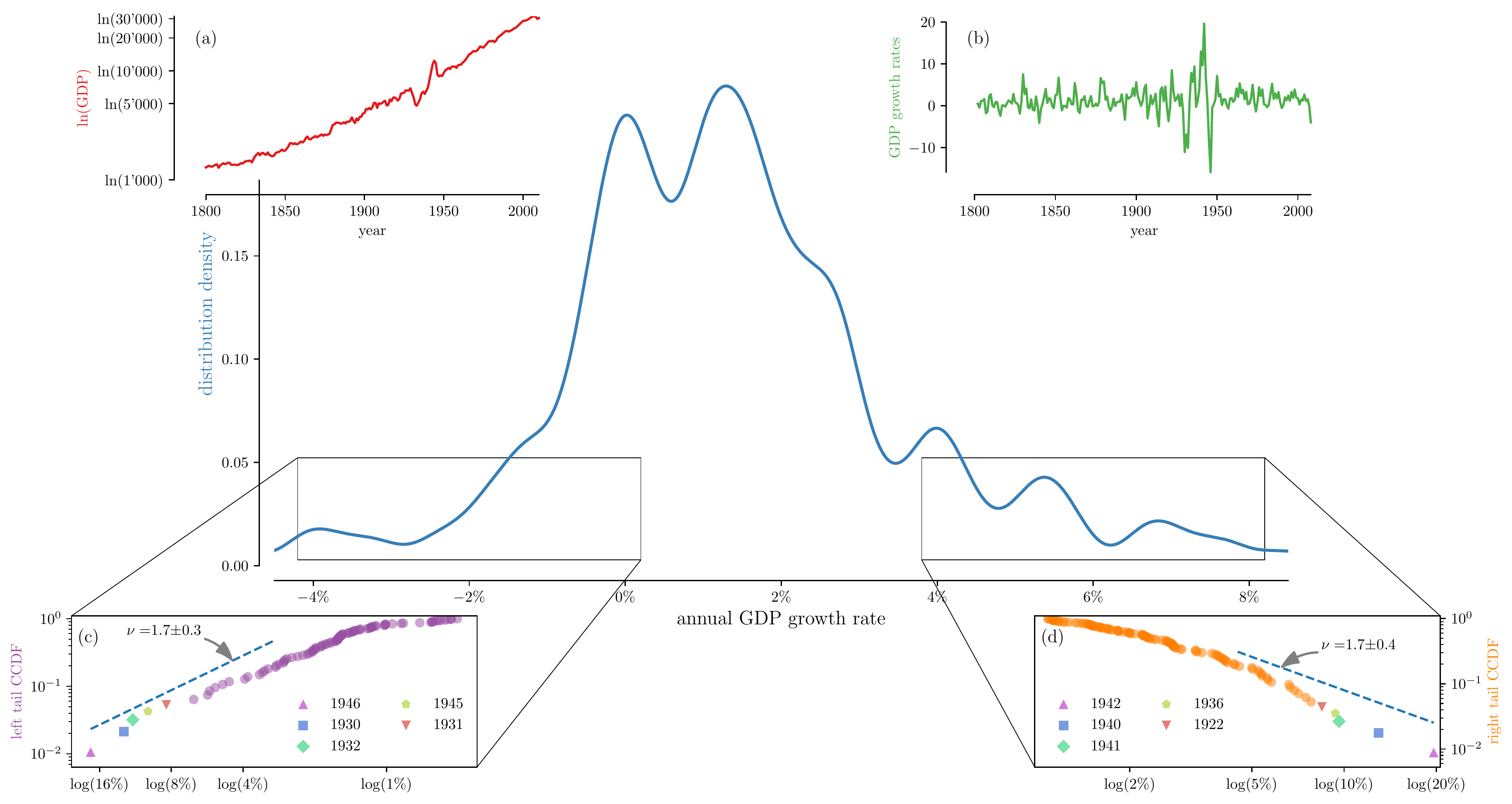}
	\caption{ 	Top left figure (a) shows the real US GDP per capita over the last 200 years. 
			Using a wavelet transform, we extract the roughly 200 annual growth rates shown in the top right figure (b). 
			The growth rate distribution is then obtained using a Gaussian kernel density estimation, shown in the main figure. 
			The bimodal structure is readily visible in the middle of the distribution. 
			The left- and right- tails of the distribution are shown in a double-logarithmic plots in figures (c) and (d), respectively. 
			The right tail is measured in absolute values, whereas the left tail is measured in terms of absolute percentage deviation from $1.5\%$, to avoid negative arguments. 
			We observe symmetric tails with CCDF tail exponents roughly between $1.5$ and $2$. 
			This is robust with respect to removal of the largest positive and negative growth rates, associated with the two world wars, the great depression and the impact on 1922 of the sharp deflationary recession of 1920-21 following the end of the WWI.}
	\label{fig:GDP_plots}
\end{figure*}

As a measure of overall economic growth, we analyze the real US GDP per capita  (from hereon, simply GDP) over the past 200 years (figure \ref{fig:GDP_plots}(a)). 
In contrast, the unnormalized total nominal GDP contains growth contributions as a result of population growth and inflation, both of which are not true sources of increased output productivity per individual. 
Given the scrutiny of the GDP in economic research, the distribution of its growth rates has also received a lot of attention. 
It is now a broadly accepted stylized fact that the distribution of GDP growth rates has tails that are fatter than Gaussian \cite{Fagiolo2008,Fagiolo2009}, and can even be power-law \cite{Fu2005,Williams2017} (but see also ref. \cite{Franke2015} for a counter argument).
Another, much less studied property is the bimodal-structure of the growth rate distribution \cite{Lera2017}.
The two peaks of the bimodal distribution can be rationalized by the fundamental out-of-equilibrium nature of the economy that is switching between boom and bust states, i.e business cycles. 

Following ref. \cite{Lera2017}, we first extract the annual GDP growth rates using a smoothening wavelet filter (figure \ref{fig:GDP_plots}(b)), 
from which we can then extract the growth rate distribution with a Gaussian density kernel estimate (figure \ref{fig:GDP_plots}, main plot). 
To the eye, the bimodal structure in the middle of the distribution is apparent. 
But one has to be careful, as this bimodal appearance is susceptible to the choice of bandwidth of the Gaussian kernel.
Statistically sound evidence for this bimodal structure has been provided in ref. \cite{Lera2017}. 

Figures \ref{fig:GDP_plots}(c) and (d) depict survival functions (CCDF) of the left and right tail of the distribution, respectively. 
The five largest positive and negative growth rates, associated with the two world wars and the great depression, are annotated explicitly.
By applying likelihood ratio tests of the power-law against the family of stretched-exponentials \cite{Malevergne2005}, we find $p$-values of $0.96$ and $0.61$ for left- and right-tail respectively.
We conclude that both tails are parsimonously represented in terms of pure power laws, without the need
for the more general parameterisation of the stretched exponential distribution family.
Using standard procedures \cite{Alstott2014}, we then determine the CCDF tail exponent and its estimated standard error. 
The exponents of left and right tail are surprisingly symmetric, and robust with respect to removal of the smallest and largest growth rates, leading merely to slightly increased error intervals.  The tail exponent $\nu$ being smaller than $2$ qualfies
the tails has belonging to the Levy-stable distribution regime \cite{GnedenkoKolmo54}.

We have thus shown that the GDP growth rate distribution exhibits a bimodal structure with symmetry power-law tails. 
In the next section, we will present a model that non-trivially combines these two apparently unrelated properties. 

\section*{L\'evy flights in steep confining potentials}
\label{sec:Levy_flight}
%%%%%%%%%%%%%%%%%%%%%%%%%%%%%%%%%%%%%%%%%%%%%%%%%%%%%%%%%%%%%%%%%%%%%%%%%%%%%%

\begin{figure*}[!htb]
	\centering
	\includegraphics[width=\textwidth]{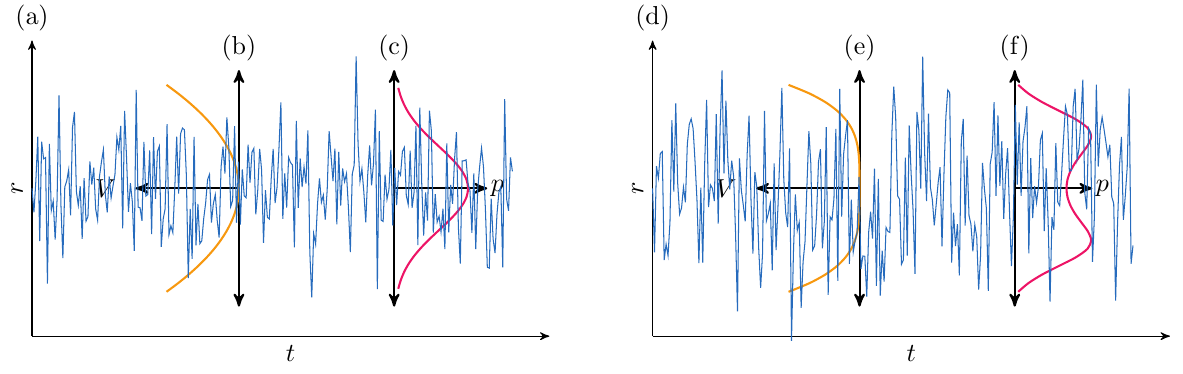}
	\caption{	Figure (a) shows the realization of an Ornstein-Uhlenbeck process \eqref{eq:generic_process} with $\beta = 2, \mu = 2, D = 1$. 
			Figure (b) depicts the quadratic potential $V(r) = \alpha r^2 / 2$ it is generated from and figure (c) is the asymptotic unimodal density distribution.
			Figure (d) shows the realization of a process \eqref{eq:generic_process} in a steep potential with
			L\'evy-stable distributed noise, with parameters $\beta = 4, \mu = 1.5, D = 1$. 
			Figure (e) depicts the corresponding quartic potential $V(r) = \alpha r^4 / 4$ and figure (f) is the asymptotic bimodal density distribution.
			 }
	\label{fig:sketch}
\end{figure*}

A generic representation of the Markovian dynamics of the GDP growth rate in continuous time is given by 
\begin{equation}
	\dd r = - \frac{ \dd V}{ \dd r}~ \dd t + \dd L(\mu,D)
	\label{eq:generic_process}
\end{equation}
where $r=r_t$ is the growth rate at time $t$ and $V$ is some function, called the `potential', which 
ensures a nonlinear mean-reversing dynamics.
The negative derivative of the potential is then interpreted as the `force' that is guiding the growth rate. 
The stochastic contribution $\dd L(\mu,D)$ denotes L\'evy-stable distributed noise with tail exponent $\mu \in (0,2]$ and scale-parameter $D > 0$. 

The most common mean-reverting model of type \eqref{eq:generic_process} is the Ornstein-Uhlenbeck process, 
characterized by a square potential $(V(r) = \alpha r^2 / 2, \alpha > 0)$ and Gaussian noise $(\mu=2)$. 
With basic stochastic methods \cite{Gardiner2009}, it can be shown that, for the Ornstein-Uhlenbeck process, the asymptotic stationary distribution of $r_t$ is Gaussian with standard deviation $D/\sqrt{2 \alpha}$. 
This is illustrated in figure \ref{fig:sketch} (left).  
In presence of Gaussian noise, the stationary distribution remains unimodal even when the square potential is replaced by a steeper potential 
\begin{equation}
	V(r) = \frac{\alpha}{\beta}~  |r-r_0|^{~\beta} ~~~~~~~ (\beta > 2).
	\label{eq:V}
\end{equation}
Here, we have also introduced a potential midpoint $r_0$ for more generality. 
Similarly, the distribution remains unimodal for a square-potential in presence of a heavy-tailed source of noise $(\mu < 2)$.
However, counter-intuitively, it has been shown \cite{Chechkin2003,Chechkin2004} that the combination of a steep potential \eqref{eq:V}, 
with heavy-tailed noise, results in a bimodal stationary distribution, as sketched in figure \ref{fig:sketch} (right). 
Furthermore, the asymptotic left and right tail of the distribution are power-laws with CCDF tail exponent $\nu \equiv \beta + \mu - 2$. 
In particular, this implies that, for $\beta > 4-\mu$, the steep potential walls confine the L\'evy noise to the extent that its tails have finite variance.

\section*{Growth rate fluctuations as confined L\'evy flights}
\label{sec:Levy_flight}
%%%%%%%%%%%%%%%%%%%%%%%%%%%%%%%%%%%%%%%%%%%%%%%%%%%%%%%%%%%%%%%%%%%%%%%%%%%%%%

We propose that the L\'evy flight in a steep potential is a good model for economic growth rates. 
Concretely, we see the L\'evy noise as a representation of the firm or sector specific idiosyncratic shocks. 
Indeed, firm size growth rates are known to be well approximated by heavy-tail distributions with infinite variance \cite{Williams2016}. 
A priori, it is not clear that fluctuations of individual firms (or sectors) would have a significant impact on aggregate measures for the economy. 
According to the central limit theorem, the average fluctuations of 
the sum of the outputs of $n$ firms decay as $\sim 1/\sqrt{n}$, a negligible effect in the limit of large $n$. 
The state of the economy is then merely influenced by economy-wide shocks such as oil price shocks, 
currency devaluation, wars, etc.
However, such arguments break down if the distribution of firm sizes is very heavy tailed \cite{Gabaix2011}, or the interdependencies of firms are asymmetric \cite{Acemoglu2012}. 
In a model that takes the microscopic firm structure into account, the individual firm shocks are propagated across the economy through a network of input-output linkages. 
To what extent the individual shocks are averaged out upon aggregation depends on the topology of the network of firm and sector dependencies \cite{Acemoglu2012}. 
If a small number of firms or sectors play a disproportionately important role as input suppliers to others, the economy is more susceptible to fluctuations of these few firms, in contrast to a more balanced scenario. 
Empirical evidence for the US \cite{Acemoglu2012}, France \cite{diGiovanni2014} and Sweden \cite{Friberg2016} suggests that indeed such asymmetries in the inter-firm and sector networks 
lead to significant contributions of a few large companies or sectors to the overall economic performance of a country. 

In our model, the microscopic firm network is represented by an effective `economic potential function'. 
The advantage of this coarse-grained perspective is that it does not rely on any specific assumptions about the underlying generating mechanisms at the micro level. 
The details average out and are captured in the potential parameter $\beta$. 
The steeper the potential, the more the individual firm shocks average out, and the less heavy the aggregate tail. 
This approach is motivated by the renormalisation group approach \cite{Wilson79}
that provides the template to represent the collective effect of many degrees of freedom at the micro level
by a few effective `renormalised' degrees of freedom at the macro level.

For more quantitative support of our proposition, we now fit the GDP data to the model. 
The L\'evy flight in the steep potential has five parameters $\alpha, \beta, r_0, \mu$ and $D$. 
Using the roughly 200 annual growth rate datapoints shown in figure \ref{fig:GDP_plots}(b), we 
calibrate these parameters with maximum likelihood, and determine the standard errors through bootstrapping. 
The fitted parameters are robust with respect to removal of individual data-points, with the exception of the two, in absolute value, largest growth rates, associated with World War II.
To obtain statistically robust estimates, we thus remove these two largest datapoints. 
Beyond statistical arguments, this is justified conceptually in the sense that a world war is a global shock that affects the economy as a whole. 
Since we are interested in contributions of individual shocks to aggregate economic output measures, 
it is justified to remove such events. 
Technically, we should thus get rid of all economy-wide shocks. 
But since only the aggregate GDP growth rates are analyzed, this is technically infeasible, and not necessary insofar as our
calibration results are not sensitive to individual datapoints, except for the WWII case.  
WWII was indeed special for the U.S, which became the factory of the world, doubling its GDP from 1940 to 1943.

\begin{figure}[!htb]
	\centering
	\includegraphics[width=0.4\textwidth]{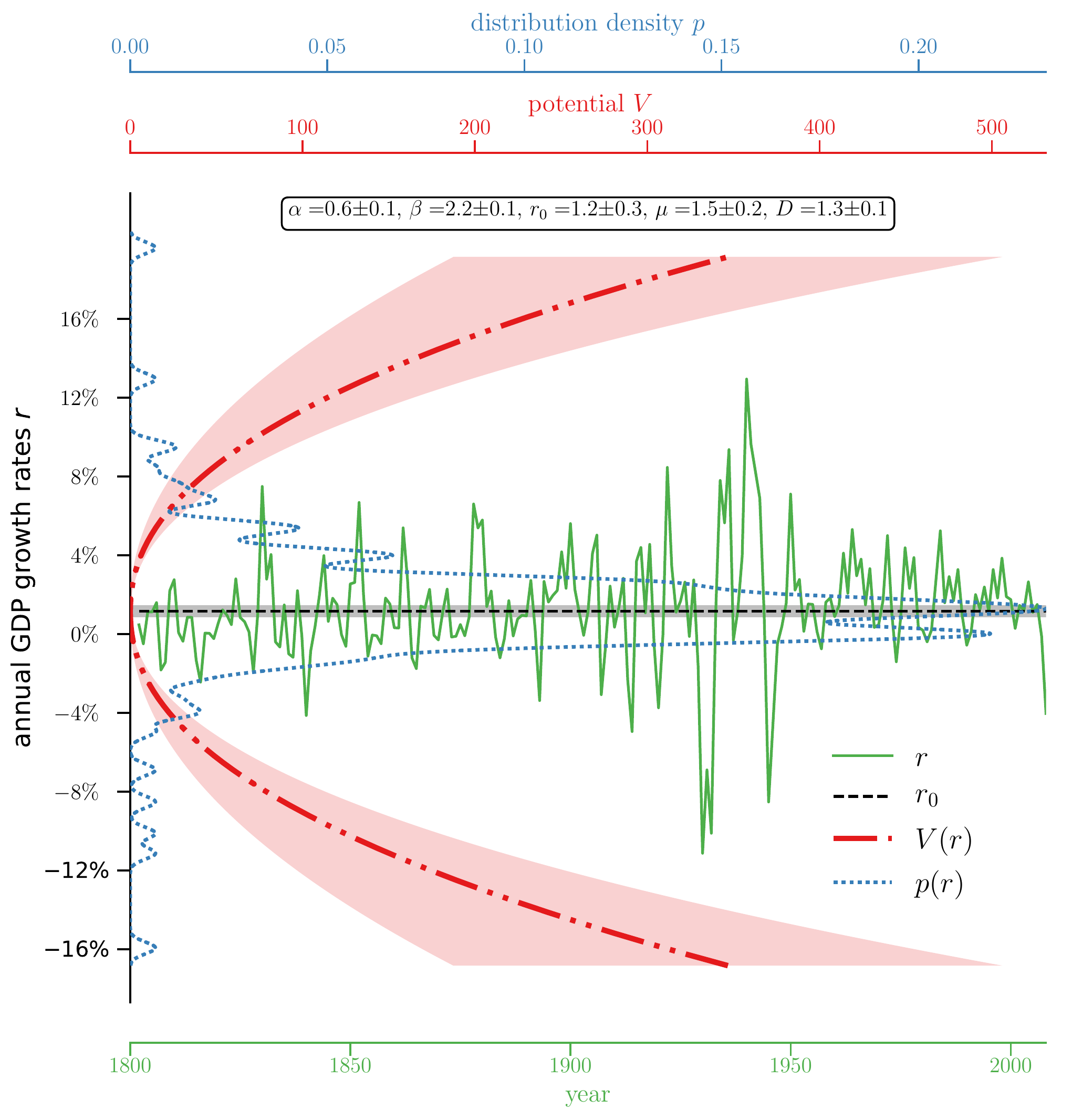}
	\caption{Using maximum likelihood, we have fitted the L\'evy flight model \eqref{eq:generic_process} in the steep potential \eqref{eq:V} to annual GDP growth rates (figure \ref{fig:GDP_plots}(b)), 
		     but ommiting the two largest growth rates in absolute value, associated with WWII. 
		    The red dotted line (and its surrounding error band) denotes the aggregate potential force. 
		    The black dashed line (and its surrounding error band) annotes the potential center-point $r_0$.}
	\label{fig:potential_fit}
\end{figure}

Details regarding the fitting procedure are found in the supporting material of this paper and the result is depicted in figure \ref{fig:potential_fit}. 
We find that the predicted GDP tail exponent $\hat{\nu} = \hat{\beta} + \hat{\mu} - 2 = 1.7 \pm 0.2$ is in excellent agreement with the actual tail exponent between $1.5$ and $2$ (figure \ref{fig:GDP_plots}(c,d)). 
The magnitude of the microscopic noise with $\hat{\mu} = 1.5 \pm 0.2$ is lighter than Cauchy noise $(\mu = 1)$, but still clearly separated from normality $(\mu = 2)$. 
The potential midpoint $\hat{r}_0 = 1.2 \pm 0.3$ is more centered towards the left of the two bimodal peaks. 
However, as additional simulations in the supplementary material confirm, this is well within the realms of expectations when analyzing just $200$ datapoints. 
We also note that the potential, with an exponent of $\hat{\beta} = 2.2 \pm 0.1$, is not much steeper than a quadratic potential with the linear mean-reverting force. 
It is just sufficiently steeper than quadratic to ensure a binomial structure of the distribution of growth rates in the presence of Levy-noise but close enough to the quadratic case
so that the GDP growth rate distribution has diverging second moment, in line with other findings \cite{Fu2005,Williams2017}. 

In conclusion, we have shown that GDP growth rates are well described as L\'evy flights in a potential that is steeper than quadratic. 
With only minimal ingredients, this model is able to capture the aggregate effect of idiosyncratic shocks to averaged economic output growth measures. 
It thereby establishes a connection between the tail exponents at a micro and macro scale, while simultaneously accounting for the bimodal structure of business cycle fluctuations.  
These are promising results that draw novel connections, and we hope that our findings will spark future research.

\section*{Acknowledgment}
\label{sec:acknowledgment}
%%%%%%%%%%%%%%%%%%%%%%%%%%%%%%%%%%%%%%%%%%%%%%%%%%%%%%%%%%%%%%%%%%%%%%%%%%%%%%
S. Lera acknowledges stimulating discussions with D. Georgiadis,  G. Britten, S. Wheatley and M. Schatz. 
The research performed by S. Lera was conducted at the Future Resilient Systems at the Singapore-ETH Centre (SEC). 
The SEC was established as a collaboration between ETH Zurich and National Research Foundation (NRF) 
Singapore (FI 370074011) under the auspices of the NRF's Campus for Research Excellence and Technological Enterprise (CREATE) programme.

\section*{References}
\balance
\bibliographystyle{unsrt}

%%%%%%%%%%%%%%%%%%%%%%%%%%%%%%%%%%%%%%%%%%%%%%%%%%%%%%%%%%%%%%%%%%%%%%%%%%%%%%

\end{document}